\RequirePackage{fix-cm}
\documentclass[twocolumn]{svjour3}          
\smartqed  
\usepackage{graphicx}
\usepackage{amsmath}
\usepackage{amssymb}
\begin{document}

\title{On Normalized Compression Distance and Large Malware 
}
\subtitle{Towards a Useful Definition of Normalized Compression Distance \\ for the Classification of Large Files}


\author{Rebecca Schuller Borbely 
}


\institute{R. Borbely \at
              CyberPoint International \\
              Baltimore, MD \\ 
              \email{rborbely@cyberpointllc.com}           
}

\date{Received: date / Accepted: date}

\maketitle

\begin{abstract}
Normalized Compression Distance (NCD) is a popular tool that uses compression algorithms to cluster and classify data in a wide range of applications.  Existing discussions of NCD's theoretical merit rely on certain theoretical properties of compression algorithms. However, we demonstrate that many popular compression algorithms don't seem to satisfy these theoretical properties. We explore the relationship between some of these properties and file size, demonstrating that this theoretical problem is actually a practical problem for classifying malware with large file sizes, and we then introduce some variants of NCD that mitigate this problem.  
\end{abstract}

\section{Introduction}
\label{intro}
In the era of big data, techniques that allow for data understanding without domain expertise enable more rapid knowledge discovery in the sciences and beyond.  One technique that holds such promise is the Normalized Compression Distance (NCD) \cite{TheSimMet}, which is a similarity measure that operates on generic file objects, without regard to their format, structure, or semantics. 

NCD approximates the Normalized Information Distance, which is universal for a broad class of similarity measures.  Specifically, the NCD measures the distance between two files via the extent to which one can be compressed given the other, and can be calculated using standard compression algorithms.   

NCD, and its open source implementation CompLearn \cite{CompLearn} have been widely applied for clustering, genealogy, and classification in a wide range of application areas.  Its creators originally demonstrated its application in genomics, virology, languages, literature, music, character recognition, and astronomy \cite{ClustComp}.  Subsequent work has applied it to plagiarism detection \cite{plagiarism}, image distinguishability \cite{image}, machine translation evaluation \cite{MT}, database entity identification \cite{database}, detection of internet worms \cite{Wehner}, malware phylogeny \cite{Wallenstein}, and malware classification {\cite{bailey} to name a few. 

Assuming some simple properties of the compression algorithm used, the NCD has been shown to be, in fact, a similarity metric  \cite{ClustComp}.    However, it remains to be seen whether real word compression algorithms actually satisfy these properties, particularly in the domain of large files.  As data storage has become more affordable, large files have become more common, and the ability to analyze them efficiently has become imperative.  Music recommendation systems work with MP3s which are typically several megabytes in size, medical images may be up to 30 MB or more \cite{medical}, and computer programs are often more than 100 MB in size.  


This paper explores the relationship between file size and the behavior of NCD, and proposes modifications to NCD to improve its performance on large files.

Section \ref{background} provides an introduction to NCD and the compression algorithm axioms that have been used for proving it to be a similarity metric.  Section \ref{large} explores the extent to which several popular (and not-so popular) compression algorithms satisfy these axioms and investigates the impact of file size on its effectiveness for malware classification.  Finally, section \ref{adapting}  proposes two possible adaptations of the NCD definition, for the purpose of improving its performance on large files, and demonstrates significant performance improvement with several compressors on a malware classification problem.

\section{NCD Background}
\label{background}

The motivating idea behind the Normalized Compression Distance is that the similarity of two objects can be measured by the ease with which one can be transformed into the other.  This notion is captured formally by the {\it information distance}, $E(X, Y)$, between two \allowbreak strings, $X$, $Y,$ which is the length of the shortest program that can compute $Y$ from $X$ or $X$ from $Y$ in some fixed programming language.  The information distance generalizes the notion of Kolmogorov complexity, where $K(X)$ is the length of the shortest program that computes $X$, and intuitively captures a very general notion of what it means for two objects to be similar. 

However, for the purposes of computing similarity, it is important that distances be relative.  Two long strings that differ in a single character should be considered more similar than two short strings that differ in a single character.  This leads to the definition of the Normalized Information Distance (NID), $$\textrm{NID}(X, Y) \equiv \frac{E(X, Y)}{\max(K(X), K(Y))}$$

The NID has several nice features:  it satisfies the conditions of a metric up to a finite additive constant, and it is universal, in the sense that it minorizes every upper semi-computable similarity distance \cite{ClustComp}.  However, it is also incomputable, which is a serious obstacle. 

Given a compression algorithm, $C$, $E(X, Y)$ can, in some sense, be approximated by $C(XY)$, the result of compressing with $C$ the file consisting of $X$ concatenated with $Y$, and $\textrm{NID}(X, Y)$ can, in turn, be approximated by $$\textrm{NCD}(X, Y) \equiv \frac{|C(XY)| - \min(|C(X)|, |C(Y)|)}{\max(|C(X)|, |C(Y)|)}$$  However, in order to prove that $NCD$ is a similarity metric, \cite{ClustComp} placed several restrictions on the compression algorithm.  A compression algorithm satisfying the conditions below is said to be a {\it normal} compressor.

\paragraph{Normal Compression}
A normal compressor, $C$, as defined in definition 3.1 in \cite{ClustComp}, is one that satisfies the following, up to an additive $O(\log n)$ term, where n is the largest length of an element involved in the (in)equality concerned:

\begin{itemize}
\item Idempotence: $|C(XX)| = |C(X)|$ and $|C(\lambda)| = 0$, where $\lambda$ is the empty string.
\item Monotonicity: $|C(XY)| \geq |C(X)|$.
\item Symmetry: $|C(XY)| = |C(YX)|$.
\item Distributivity: \\$|C(XY)| + |C(Z)| \leq |C(XZ)| + |C(YZ)|$. 
\end{itemize}
where $C(X)$ denotes the string $X'$ resulting from the application of compressor $C$ to string $X,$ $XY$ denotes the concatenation of $X$ and $Y,$ and $|X|$ denotes the length of string (or file) $X$.

The question remains whether existing compression algorithms satisfy these axioms, particularly in the domain of large files.  While NCD has apparently been quite successful in practice, the majority of applications have been on relatively small files.  (See section \ref{intro}.)  Notably, music applications \cite{music,ClustComp}, used MIDI files rather than the more common, and much larger, MP3 format.  

Previous work \cite{Pitfalls} explored the NCD distance from a file to itself (which is closely related to the idempotence axiom) for bzip, zlib, and PPMZ on the Calgary Corpus \cite{Calgary}, comprising 14 files, the largest of which is under 1 MB.   The following section explores these axioms on a larger and more representative dataset and investigates the practical impact of deviations from normality.

\section{Application of NCD to Large Files}
\label{large}

\subsection{Normality of Compression Algorithms}

The definition of a normal compressor deals with asymptotic behavior, allowing for an $O(\log(n))$ discrepancy in the axioms of idempotence, monotonicity, symmetry, and distributivity.  Thus, in theory, experimental validation (or refutation) of these axioms is not truly feasible -- perhaps the behavior changes when the file size is beyond that of the largest file in our experiment.  Nonetheless, we endeavor to experimentally explore these axioms more extensively than has been done in prior work.

\paragraph{Data}
We combined the traditional Calgary Corpus with the Large and Standard Canterbury Corpora, as well as the Silesia Corpus\footnote{These are standard corpora for the evaluation of compression algorithms and are available at http://www.data-compression.info/Corpora/}.  The latter contains files of size ranging from 6 MB to 51 MB, greatly expanding the size distribution over the previous corpora.  

\paragraph{Idempotence} 

\begin{figure*}
\includegraphics[width=0.75\textwidth]{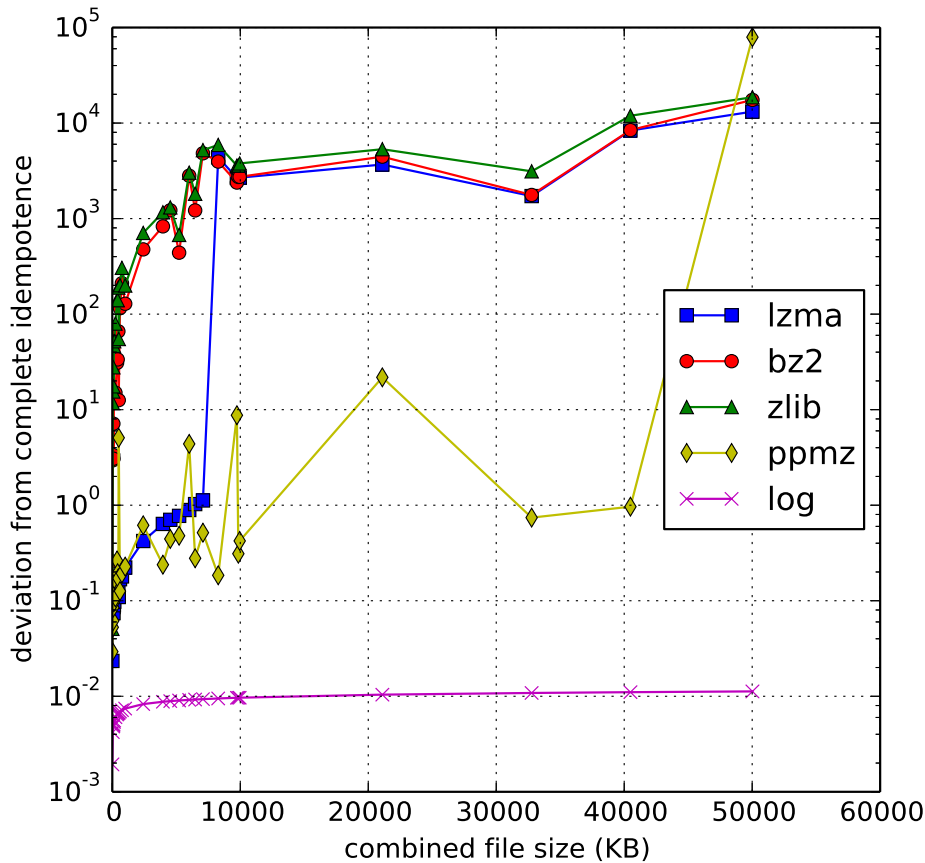}
\caption{Idempotence on compression corpora: $|C(XX)|-|C(X)|$ as compared to $\log(|XX|)$ versus $|XX|$. }
\label{Idempotence}
\end{figure*}

\begin{figure*}
\includegraphics[width=0.75\textwidth]{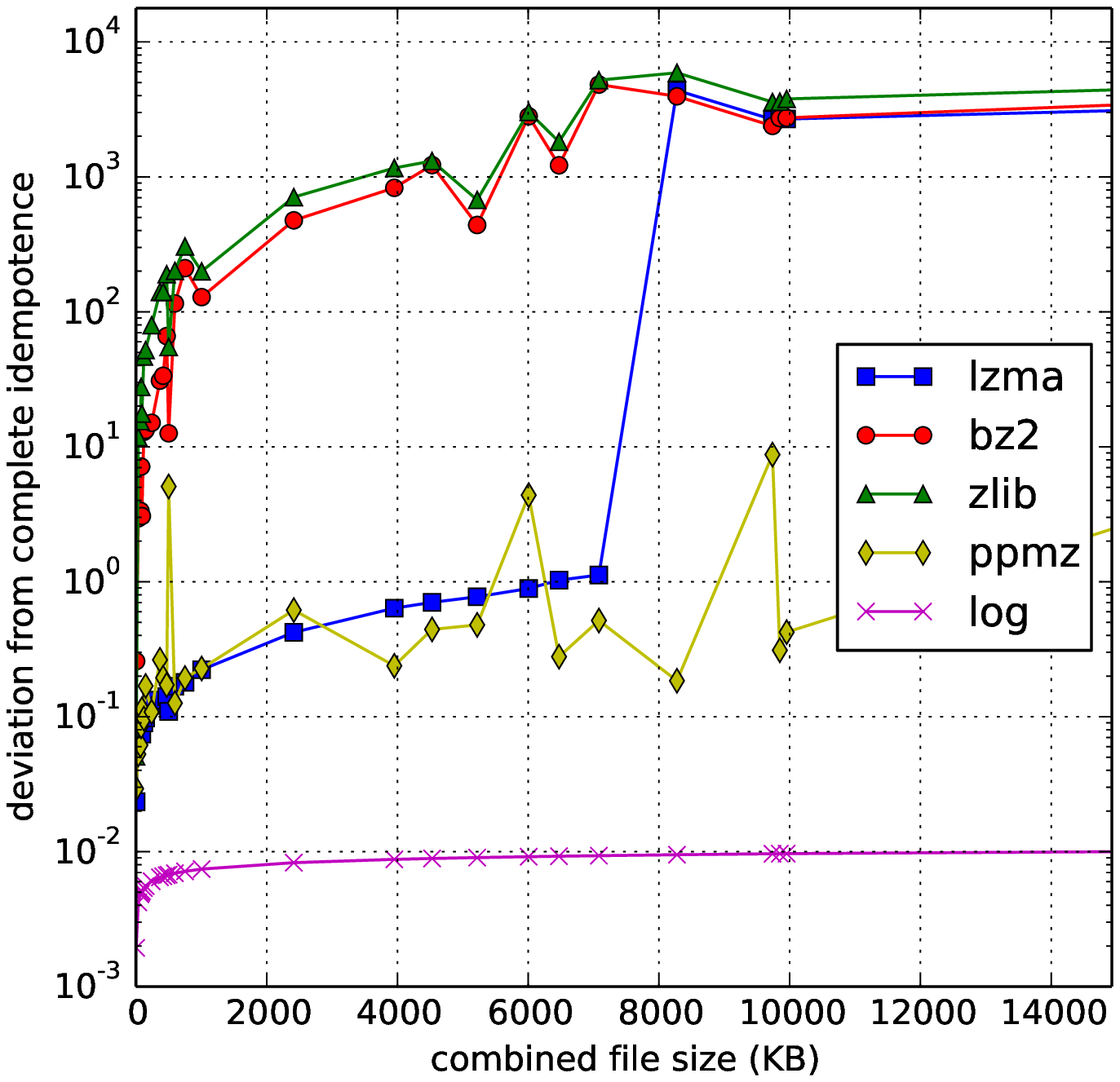}
\caption{Idempotence on compression corpora: Enlargement of a portion of the graph in figure \ref{Idempotence} to more clearly show the behavior for smaller files. }
 \label{IdempotenceZoom}
\end{figure*}

Figures \ref{Idempotence} and \ref{IdempotenceZoom},  show the difference in the sizes of $C(X)$ and $C(XX)$, and $\log(|XX|)$, for a representative subset of files $X$ in the dataset, with $C$ ranging over compression algorithms bzip2 \cite{bzip2}, lzma \cite{lzma}, PPMZ \cite{ppmz}, and zlib \cite{zlib}.  Indeed, bz2 and zlib quite apparently fail the idempotence axiom, with $|C(XX)|$ growing much faster than $|C(X)|$, with a factor of \allowbreak $\log(|XX|)$ unable to put a dent in the difference.  While PPMZ and lzma appear significantly better, still, this value grows much faster than $\log(|XX|)$, as apparent in figure \ref{IdempotenceZoom}.  We see that lzma makes a large jump around 8 MB (but even before that, its growth is much larger than the $\log$ function).

\paragraph{Symmetry}

Figure \ref{Symmetry} shows the magnitude of difference between $|C(XY)|$ and $|C(YX)|$. While in most cases, at this scale, this was bounded by $\log(|XY|)$ (and in all cases by a small constant factor thereof), the asymptotic behavior is unclear, as values for all four compressors spike wildly.  This is likely due to the fact that the extent of the symmetry is highly dependent on the compressibility and similarity of the two files involved.  zlib and lzma look quite promising for symmetry, while the asymptotic behavior of PPMZ and bz2 is not discernible.  

\begin{figure*}
\includegraphics[width=0.75\textwidth]{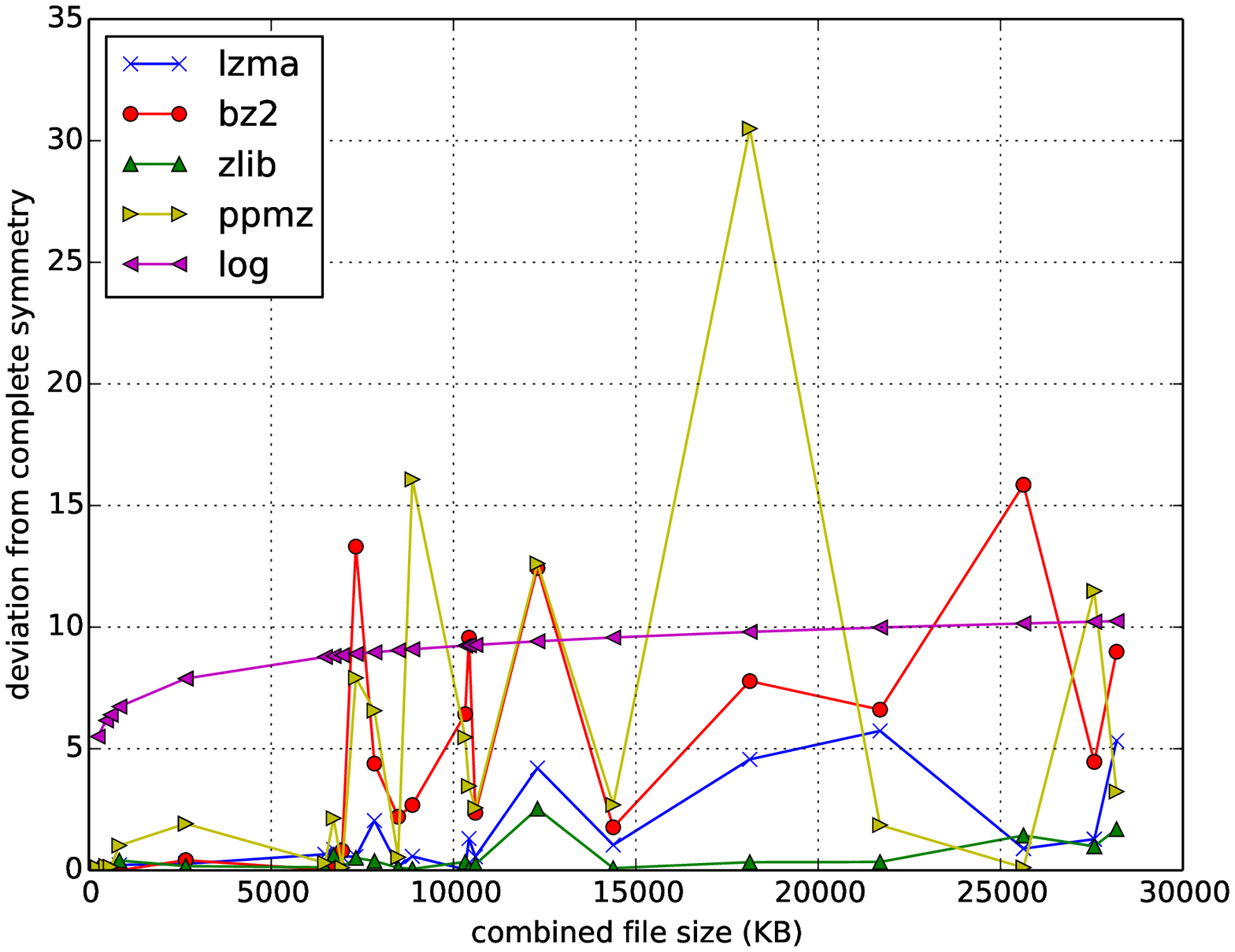}
\caption{Symmetry: The difference between $|C(XY)|$ and $|C(YX)|$, as compared to $\log(|XY|).$}
 \label{Symmetry}
\end{figure*}

\paragraph{Distributivity and Monotonicity}
Initial experiments with distributivity and monotonicity did not give cause for concern.

\paragraph{} Our experiments have shown serious violation of the idempotence axiom that has been used to prove theoretical properties of NCD, leaving a potential gap between theory and practice.  The next section explores the extent to which NCD can be useful in spite of this gap.

\subsection{Classification using NCD with Abnormal Compressors}\label{class}

We have demonstrated that none of the compression algorithms we explored satisfy the requirements for normal compression.  The question remains whether this contraindicates their use with NCD.  As mentioned above, much previous work has demonstrated NCD's utility with some of these compression algorithms in applications with small file sizes.  However, the compressors' deviation from normality grows with file size.  Do they remain useful with with larger files?

To address this question, we explored the accuracy of NCD in identifying the malware family of APK files from the Android Malware Genome Project dataset \cite{NCSU,NCSUpaper}.  In particular, we took a subset of 500 samples from the Geinimi, DroidKungFu3, DroidKungFu4, and GoldDream families.\footnote{We selected these families due to their containing enough samples to allow for a meaningful test, and containing large enough files to challenge the compressors.}  Geinimi samples in this dataset have size up to 14.1 MB, DroidKungFu3 up to 15.4 MB, DroidKungFu4 up to 11.2 MB,  and GoldDream up to 6.4 MB.  

We evaluated the NCD with the same four compression algorithms as above, using a nearest neighbor classifier \cite{KNN} with a single (randomly selected) instance of each malware family in the reference set.\footnote{For readers unfamiliar with nearest neighbor classification, specifically we classified a "test" sample by looking at the distance between it and each of the "reference" samples, and selecting the family of the nearest (i.e. most similar) reference sample.} Note that we intentionally restricted the reference set to make the classification problem difficult in order to explore the limitations of the compression algorithms when used with NCD.  Results are shown in figure \ref{Basic_NCD_Android}. In spite of clearly violating the idempotence property, both lzma and PPMZ performed significantly better than random guessing.  In line with their relative normality, lzma performed best, at, 59.7\%  with PPMZ up next at 44.4\%.   Although bz2 is slightly closer to satisfying the idempotence property than zlib, zlib actually outperformed bz2, albeit not by much, with accuracies of 33.3\% and 29.8\%, respectively, with neither performing much better than random guessing.   

\begin{figure}
\includegraphics[width=0.45\textwidth]{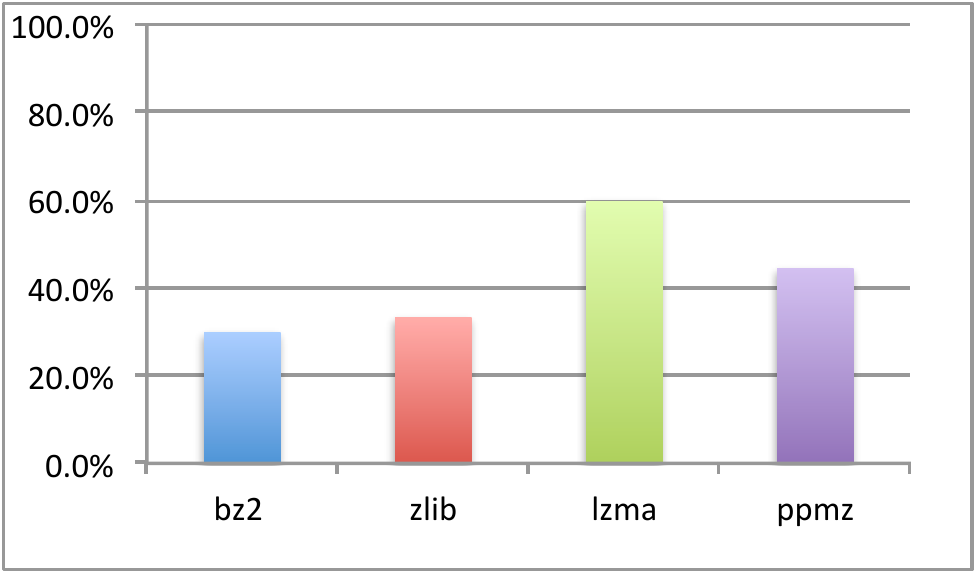}
\caption{Accuracy of NCD in identifying Android malware family, using a 1-NN classifier}
 \label{Basic_NCD_Android}
\end{figure}

To demonstrate the relevance of file size, we performed the same test with one slight change, this time using only reference samples smaller than 200 KB.
We saw drastic improvement with bz2 (now 75.4\%), lzma (82.5\%), and PPMZ (66.7\%), while zlib's performance actually got worse (29.2\%). 

Finally, looking only at files smaller than 200 KB yielded improved performance by bz2 (89.7\%), zlib \allowbreak (37.9\%), and PPMZ (75.9\%), but lzma actually performed slightly worse (75.9\%).   The latter suggests that file size is not the only factor that can inhibit the performance of a compression algorithm with NCD.  Notably, bz2 outperformed lzma on these files.  These results are shown in figure \ref{Basic_NCD_sizes}.

\begin{figure*}
\includegraphics[width=0.75\textwidth]{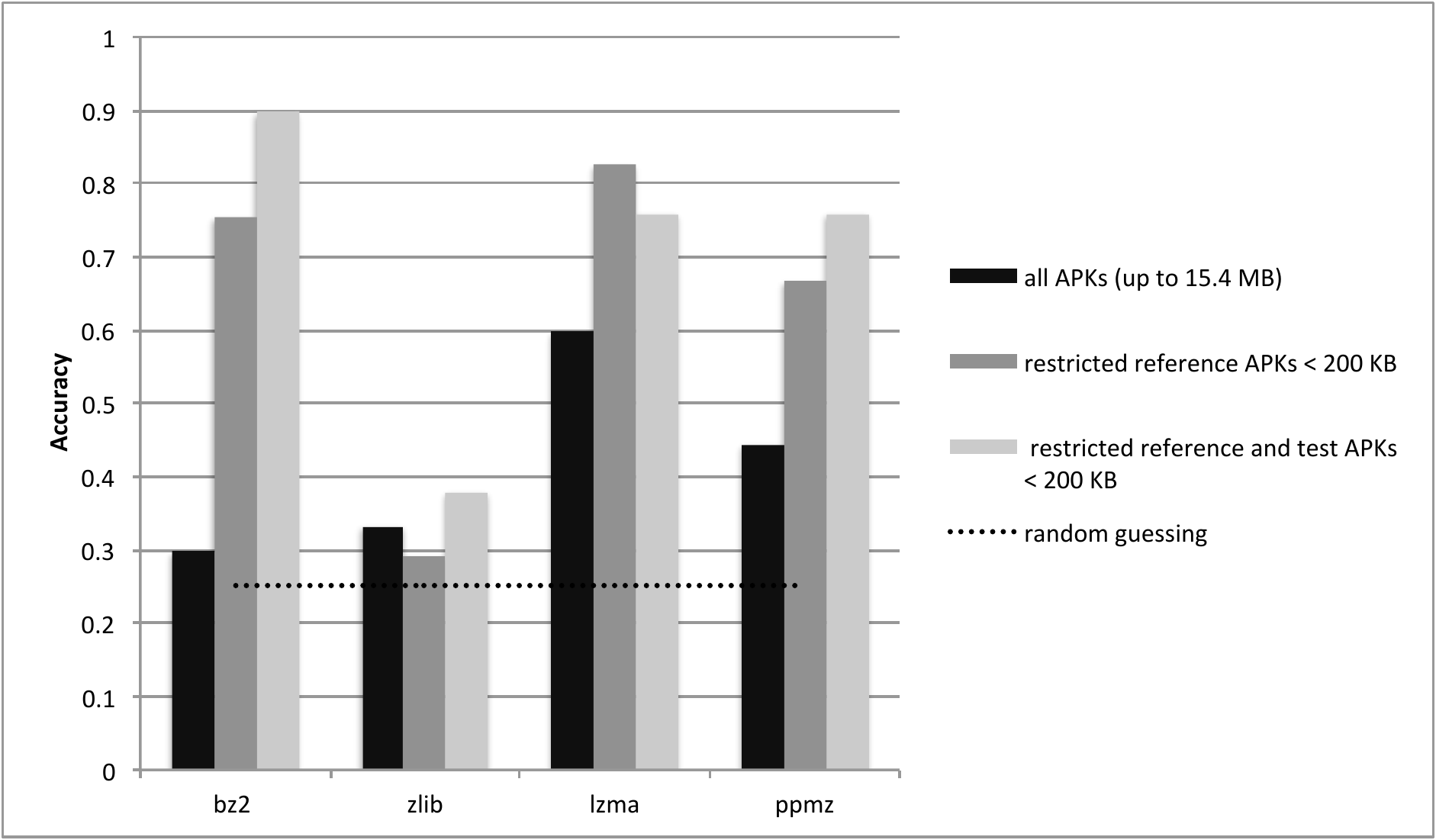}
\caption{Effect of file size on accuracy of NCD in identifying Android malware family, using a 1-NN classifier}
 \label{Basic_NCD_sizes}
\end{figure*}

\section{Adapting NCD to Handle Large Files}
\label{adapting}

We saw in section \ref{class} that NCD has widely varying performance on large files, depending on the compression algorithm used.  The memory limitations of the algorithm are key here.  The major hurdle is to effectively use information from string $X$ for the compression of string $Y$ in computing $C(XY)$.   Algorithms like bz2 and zlib have an explicit block size as a limiting factor; if $|X| > \textrm{block\_size}$, then there is no hope of benefiting from any similarity between $X$ and $Y$.   In contrast, lzma doesn't have a block size limitation, but instead has a finite dictionary size; as it processes its input, the dictionary grows.  Once the dictionary is full, it is erased and the algorithm starts with an empty dictionary at whatever point it has reached in its input.  Again, if this occurs before reaching the start of $Y$, hope of detecting any similarity between $X$ and $Y$ is lost.  Likewise, even if $X$ is small, but $Y$ is large, with the portion of $Y$ that is similar to $X$ appearing well into $Y$, the similarity can't be detected.  

Thus, it seems logical that we could improve the effectiveness of NCD by bringing similar parts of $X$ and $Y$ in closer proximity of one another; rather than computing NCD using $C(XY)$, we propose using $C(J(X, Y))$ where $J$ is some method of combining strings X and Y.  So, we define $$NCD_{C, J} = \frac{|C(J(X, Y))| - min(|C(X)|, |C(Y)|)}{max(|C(X)|, |C(Y)|)}.$$ In the original definition of NCD, $J$ is simply concatenation.  In an ideal world, $J$ would locate similar chunks of $X$ and $Y$ and place them adjacently.  However, if $J$ is too destructive of the original strings, much of the original compression of $X$ and $Y$ individually will be lost, resulting in a higher overall value for $NCD_{C, J}(X, Y)$.  Thus, we want these similar chunks to be as large as possible so as to still allow both chunks to fit within the block size, or to allow processing of them both within the same dictionary.   There are some simple ways to achieve this.

One approach would be to apply a string alignment algorithm to $X$ and $Y$, and combine the two strings so that aligned segments are located in sufficient proximity.  However, while Hirschberg's algorithm \cite{Hirschberg} allows for such alignment to be performed in linear space, thus eliminating memory issues, it takes time proportional to the product of the file sizes and is thus quite slow with large files.  Further, this is limited to finding a very specific type of similarity, which is order-dependent.  However, we propose two other approaches inspired by this notion.

\paragraph{\textbf{Interleaving}} The simplest approach is to assume that similar parts of $x$ and $y$ are similarly located, and just weave them together in chunks of size $b$.  Say $X=x_1x_2...x_n$ and $Y=y_1y_2...y_m$, where $|x_i| = |y_j| = b$ for $1\leq i \leq n-1 $ and $1 \leq j \leq m-1$, $0 \leq |x_n| < b$, and $0 \leq |y_m| < b$.  Then define $$J_b(x, y) = \begin{cases} 
	\hfill  x_1y_1x_2y_2 \dots x_ny_ny_{n+1}...y_m \hfill  & \text{if $n<m$} \\
	\hfill  x_1y_1x_2y_2 \dots x_my_mx_{m+1}...x_n \hfill  & \text{otherwise} \\
\end{cases}$$

\paragraph{\textbf{NCD-shuffle}} Another approach is to split both strings into chunks of the desired size (selected to be appropriate for the compression algorithm) and apply the traditional NCD to determine the similarity of each chunk of $X$ to each chunk of $Y$, and align them accordingly, with the most similar chunks from the two strings adjacent.  

\subsection{NCD Adaptation Results}

Using the original classification problem from section \ref{class}, we applied the interleaving (IL) and NCD-shuffle (NS) file combination techniques with various block sizes with each of the compression algorithms.  As shown in table \ref{Android_heuristics} and figure \ref{AndroidImprovement}, in all cases, one or both techniques yielded a better performance than the traditional NCD.  Figure \ref{AndroidImprovement} also includes the accuracy when 5 representatives from each family are used for comparison (with the exclusion of PPMZ, which was too slow for this experiment).  Most notably, these techniques boosted bz2 from 29.8\% accuracy to 52.2\% accuracy with a single training sample, and from 55.2\% to 75.2\% with 5 training samples, and boosted zlib from 30\% to 74.8\% with 5 training samples.  

\begin{table}
\caption{Comparison of performance of different combining functions with NCD in a 1-NN classifier for Android malware family identification, with varying block sizes (block sizes in thousands of KB)}
\begin{tabular}{l l l l l l l}
\hline\noalign{\smallskip}
 &	concat	& IL 1	& IL 10	& IL 100	& IL 1000	\\ 
 \noalign{\smallskip}\hline\noalign{\smallskip}

bz2		& 0.298	& 0.464	& 0.462	& 0.456	& 0.308	\\
zlib		&0.333&	0.19 &	0.194&	0.131&	0.317&	\\
lzma		&0.597	&0.637&	\textbf{0.643} &	0.635	&0.603	 \\
PPMZ	&0.444&	0.357	&\textbf{0.484}	&0.438&	0.442	\\
\noalign{\smallskip}\hline
\end{tabular} 
\smallskip
\begin{tabular}{ l l l l l l }
\hline\noalign{\smallskip}
 &	concat	&  NS 10& NS 100& NS 1000 \\
 \noalign{\smallskip}\hline\noalign{\smallskip}

bz2		& 0.298		& \textbf{0.522}	& 0.423	& 0.325 \\
zlib		&0.333&		\textbf{0.433}	&0.200	& 0.325 \\
lzma		&0.597		&0.641&	\textbf{0.643}&	0.627 \\
PPMZ	&0.444		&0.371&	0.438&	0.435 \\
\noalign{\smallskip}\hline
\end{tabular}
\label{Android_heuristics}
\end{table}

\begin{figure*}
\includegraphics[width=.75\textwidth]{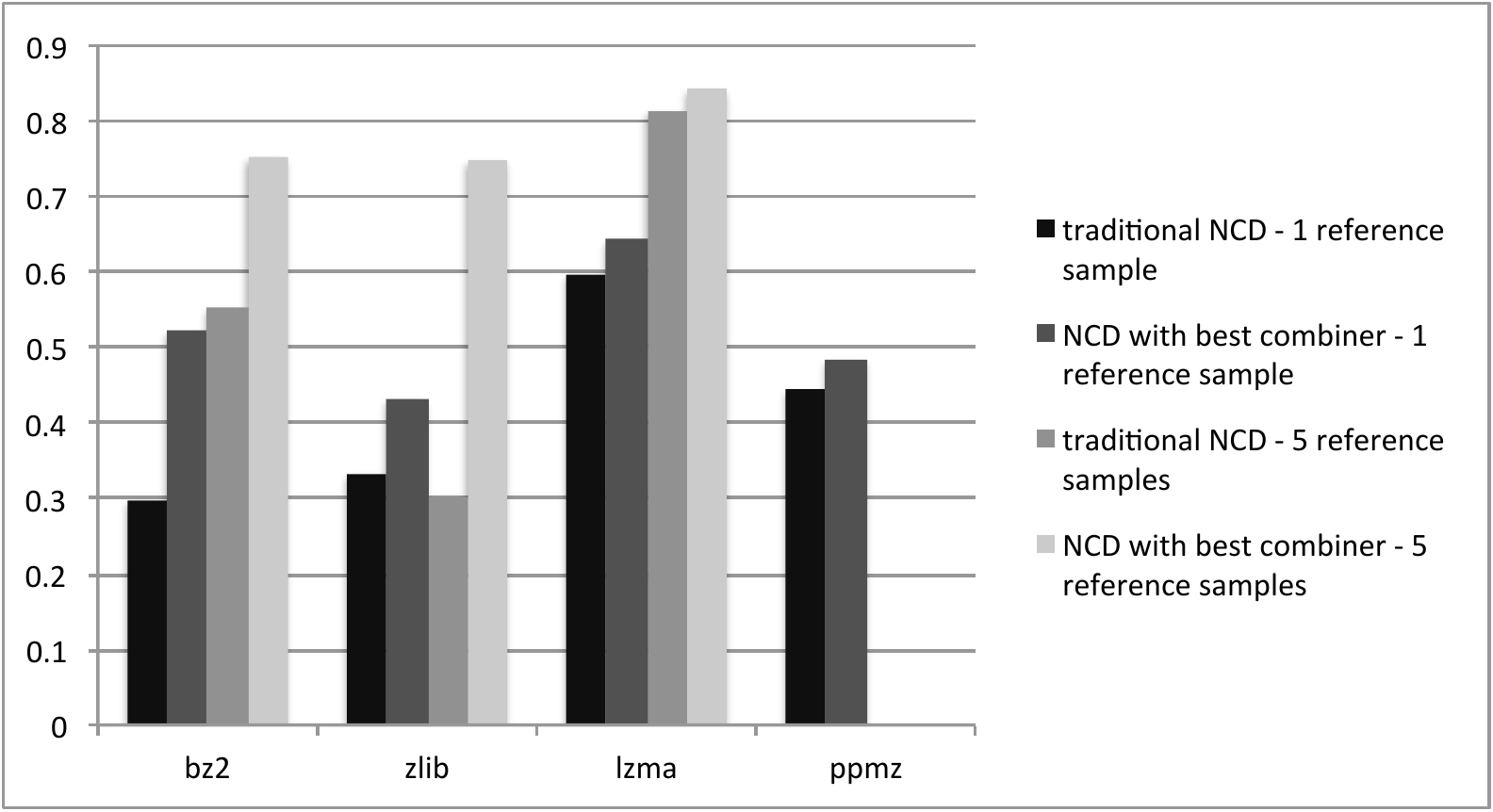}
\caption{Traditional NCD compared to the best of the alternative combiners we explored for Android malware family identification}
 \label{AndroidImprovement}
\end{figure*}

Note that we also performed smaller experiments on music MP3 data and medical image data, and also saw improvements there\footnote{For example, on a set of 66 mammography images from DDSM \cite{DDSM,DDSMcurrent}, zlib improved from 31.3\% accuracy to 54.7\% accuracy in identifying cancerous images, and bz2 improved from 26.6\% to 62.5\% accuracy.}, so we expect these techniques to offer improvement not just in malware classification, but in all domains where large files are prevalent.

\section{Conclusion and Future Directions}
We have demonstrated that several compression algorithms, lzma, bz2, zlib, and PPMZ, apparently fail to satisfy the properties of a normal compressor, and explored the implications of this on their capabilities for classifying Android malware with NCD.   More generally, we have shown that file size is a factor that hampers the performance of NCD with these compression algorithms. Specifically, we found that lzma performs best on this classification task when files are large (at least in the range we explored), but that bz2 performs best when files are sufficiently small.  We have also found zlib to generally not be useful for this task.  PPMZ, in spite of being the top performer in terms of idempotence,  did not come close to the most accurate compressor in any case.

We introduced two simple file combination techniques that boost the performance of NCD on large files with each of these compression algorithms.  

However, the challenges of choosing the optimal compression algorithm and the optimal combination technique (and parameters therefor) remain.  For supervised classification applications, it is easy enough to use a test set to aid in the selection of the technique and block size parameter for the relevant domain.  However, for clustering or genealogy tasks, the burden remains to study several resulting clusterings or hierarchies to determine which is most appropriate.  

It remains for future work to better understand what properties of a data set make it more or less amenable to the different compression algorithms and different combination techniques and parameters.

Nonetheless, these techniques offer enhanced NCD performance in malware classification (as well as other tasks) with large files, and suggest that further research in this direction is worth pursuing.  


\bibliographystyle{spmpsci}      
\bibliography{TowardsNCD}   

\end{document}